\documentclass[twocolumn,superscriptaddress,aps,prl]{revtex4}
\usepackage{graphicx}
\usepackage[novolumeabbr]{unitsdef}

\begin{document}
\bibliographystyle{prsty}

\title{Temperature-dependent terahertz conductivity of topological insulator Bi$_{1.5}$Sb$_{0.5}$Te$_{1.8}$Se$_{1.2}$}
\author{Chi Sin Tang}
\author{Bin Xia}
\author{Xingquan Zhou}
\affiliation{Division of Physics and Applied Physics, School of Physical and Mathematical Sciences, Nanyang Technological
University, Singapore 637371, Singapore}
\author{Jian-Xin Zhu}
\affiliation{Theoretical Division, Los Alamos National Laboratory, Los Alamos NM 87545, USA}
\author{Lan Wang}
\author{Elbert E. M. Chia}
\affiliation{Division of Physics and Applied Physics, School of Physical and Mathematical Sciences, Nanyang Technological
University, Singapore 637371, Singapore}

\date{\today}

\begin{abstract}
Using Terahertz Time-Domain Spectroscopy, we study the temperature-dependent complex optical conductivity of the topological insulator, Bi$_{1.5}$Sb$_{0.5}$Te$_{1.8}$Se$_{1.2}$ single-crystal from 5~K to 150~K in the terahertz regime (0.4 -- 3.0~THz). We analyze our experimental results using the Drude-Lorentz model, with the Drude component representing the metallic surface state and the Lorentz term representing the bulk insulating state. We find the conductivity to be dominated by the Drude contribution, suggesting the presence of metallic surface states. The low-frequency real conductivity follows a thermally-activated behavior. Its origin is also discussed.
\end{abstract}

\maketitle

Topological insulators are electronic materials with an insulating bulk and conducting surface \cite{Moore2010,Hasan2010}. Such unique features have been attributed to the presence of time-reversal symmetry and spin-orbit interaction \cite{Hsieh2008,Hasan2010}. Bi$_{2}$Se$_{3}$, Bi$_{2}$Te$_{3}$ and Sb$_{2}$Te$_{3}$ are several of the semi-metallic chalcogenides shown to have the properties of a three-dimensional topological insulator \cite{Zhang09,Xia09,QiStanford2010}. Investigations of the surface state have been made by surface probes such as Angle-Resolved Photoemission Spectroscopy (ARPES) \cite{Hsieh09,Hsieh2008,Roushan2009} and Scanning Tunneling Microscopy (STM) \cite{Zhang09,Alpichshev10}. In Bi$_{2}$Se$_{3}$ and Bi$_{2}$Te$_{3}$, the presence of non-stiochiometry-induced bulk carriers induced extrinsic metallic conduction in the bulk \cite{Checkelsky2009,Eto2010,Peng2010,Analytis2010a}, making it difficult to detect the metallic surface states. Recently, new topological insulators Bi$_{2}$Te$_{2}$Se and Bi$_{1.5}$Sb$_{0.5}$Te$_{1.7}$Se$_{1.3}$ were found \cite{Ren10,Taskin11}, where non-stiochiometry-induced donors and acceptors compensate each other, yielding high bulk resistivities of 1--10 $\Omega$.cm and high contribution ($\sim$70$\%$) of the total conductance by surface transport, thus showing a greater contrast between the bulk and surface resistivities compared to Bi$_{2}$Se$_{3}$, Bi$_{2}$Te$_{3}$ and even Bi$_{2}$Te$_{2}$Se \cite{Pietro12}. ARPES experiments have also shown that their chemical potential is always located inside the bulk band gap \cite{Arakane12}.

Terahertz Time-Domain Spectroscopy (THz-TDS) is a non-contact far-infrared optical technique suitable for probing the low-energy excitations of strongly correlated electronic systems such as cuprate superconductors \cite{Averitt02}, pnictide superconductors \cite{RVAguilar2010}, and colossal magnetoresistance manganites \cite{Averitt02}. Recently, Aguilar \textit{et al.} \cite{RVAguilar2012a,RVAguilar2012b} applied the same technique to study the THz response of topological insulator Bi$_{2}$Se$_{3}$ thin films, and had to account for free carriers from the bulk in characterizing the conductivity of the material. In this Paper, we present THz-TDS studies of Bi$_{1.5}$Sb$_{0.5}$Te$_{1.8}$Se$_{1.2}$ (BSTS) single crystal at temperatures 5~K--150~K, to study the frequency and temperature-dependent optical conductivity in the far-infrared regime (0.4 -- 3.0~THz). By modeling the total conductance as the sum of Drude (surface) and Lorentz (bulk) components, we obtained the surface conductivity $\sigma_{surf}\sim$10$^{5}$~($\Omega$.cm)$^{-1}$ to be 10$^{4}$ times the bulk conductivity $\sigma_{bulk}\sim$10$^{1}$~($\Omega$.cm)$^{-1}$. This suggests a significant contribution of the topological surface states to the conductivity of the BSTS sample.

Bi$_{1.5}$Sb$_{0.5}$Te$_{1.8}$Se$_{1.2}$ single crystals were synthesized by melting high-purity ($99.9999\%$) Bi, Sb, Te and Se with molar ratio 1.5:0.5:1.8:1.2 at $850^{\circ}\text{C}$ in an evacuated quartz tube. The temperature was then gradually decreased to room temperature over a span of three weeks \cite{Xia2011}. The BSTS single crystal was then cleaved repeatedly to a thickness of $\sim$60~$\mu$m while maintaining a large surface area ($\sim$4$\times$4~mm$^{2}$).

THz transmission of the BSTS single crystal were measured using a conventional THz-TDS system (TeraView Spectra 3000) incorporated with a Janis ST-100-FTIR cryostat. The THz signal was generated and detected by photoconductive antennae fabricated on low-temperature-grown GaAs films. The aperture diameter is $3.5$~mm, allowing for an accurate measurement of the THz signal down to $\sim$0.4~THz. The time-domain electric field of the THz pulse signal is transmitted through the BSTS sample ($\vec{E}_{S}(t)$), while the reference signal ($\vec{E}_{R}(t)$) is transmitted through vacuum. 1800 THz traces were taken in 60 seconds for each reference or sample run. The sample holder was moved back and forth between the sample and reference positions by means of a vertical motorized stage with a resolution of 2.5~$\mu\text{m}$. Fast Fourier Transform (FFT) was then performed on the time-domain THz signal to obtain the amplitude and phase of the THz spectra. Since the THz-TDS detects both the amplitude and phase of the THz signal, there is no need to use the Kramers-Kronig transformation to extract the real and imaginary components of the material optical parameters.

\begin{figure}\centering
\includegraphics[width=8cm,clip]{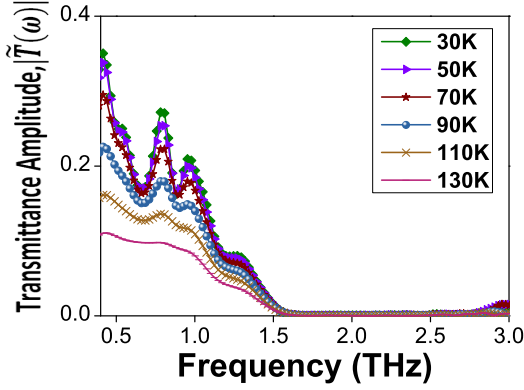}
\caption{Amplitude of complex transmittance, $|\tilde{T}(\omega)|$, at various temperatures.}
\label{fig:Transmittance_All}
\end{figure}

By finding the ratio between the sample [$\tilde{E}_{S}(\omega)$] and reference [$\tilde{E}_{R}(\omega)$] spectra, one is able to obtain the complex transmittance, $\tilde{T}(\omega)$=$\tilde{E}_{S}(\omega)/\tilde{E}_{R}(\omega)$. Figure~\ref{fig:Transmittance_All} shows the amplitude of the experimental transmittance of BSTS, $|\tilde{T}(\omega)|$, at various temperatures. A series of wiggles have been observed at low temperatures which eventually get washed out with increasing temperature above $\sim$100~K. We associate these wiggles to the presence of impurity states in our BSTS sample due to doping. Similar wiggles at low temperatures were also seen in optical conductivity data of Bi$_{2}$Se$_{3}$, Bi$_{2}$Se$_{2}$Te, which the authors associated with hydrogen-like 1\textit{s}$\rightarrow$\textit{np} transitions of isolated impurities \cite{Pietro12}. In addition, a drop in $|\tilde{T}(\omega)|$ to zero has been observed at $\sim$1.7~THz and after which, rises again at $\sim$2.8~THz. This corresponds to an optical phonon mode at 1.9~THz, to be described later. By fitting the experimental $\tilde{T}(\omega)$ to the theoretical expression
\begin{multline}
\tilde{T}(\omega) = \frac{4 \tilde{n}  \exp[i\omega d(\tilde{n} - 1) / c ]}{(1 + \tilde{n})^{2} - (1 - \tilde{n})^{2}\exp[ 2 i\omega d \tilde{n} / c ]}\\
\times \sum_{k=0}^{M}\biggl[\biggl(\frac{1 - \tilde{n}}{1 + \tilde{n}}\biggr)^{2}\exp{\biggl(2i\frac{\tilde{n}\omega d}{c} \biggr)}\biggr]^{M}
\label{Transmittance}
\end{multline} one is able to able to obtain the complex refractive index of the sample, $\tilde{n} = n + i\kappa$. Here $d$ (= 60~$\mu$m) is the sample thickness, $c$ is the speed of light in vacuum, and $M$=6 is the number of multiple reflections of the THz pulse within the sample. Note that the wiggles seen below 80~K are not a result of multiple reflections within the sample --- their existence is robust when we vary $M$ around its expected value.

\begin{figure}\centering
\includegraphics[width=8cm,clip]{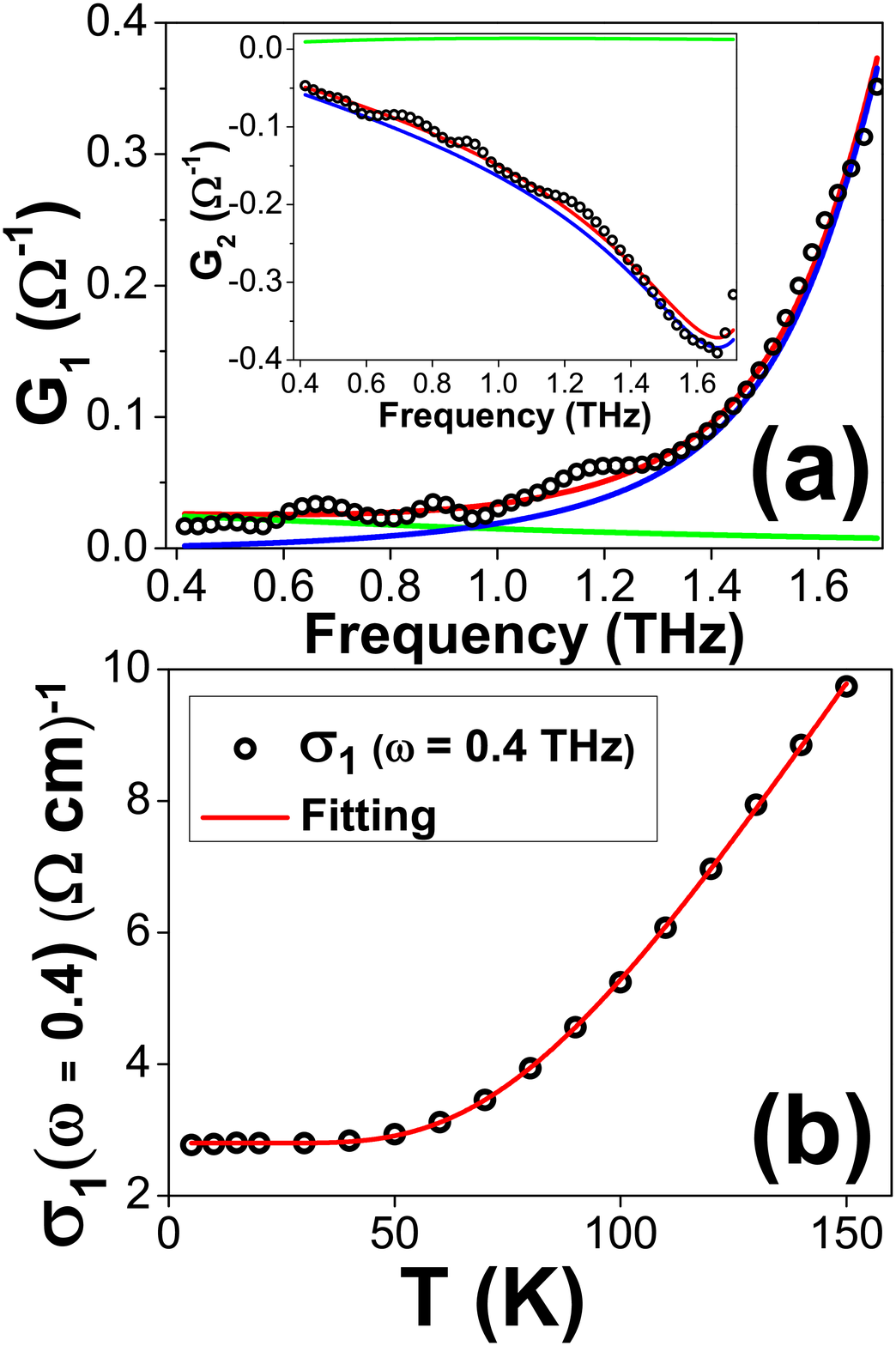}
\caption{(a) Real conductance $G_{1}(\omega)$ at 30~K. Inset shows the imaginary conductance $G_{2}(\omega)$. (O) = data. Solid lines = Drude (green) and Lorentz (blue) contributions, and the fit to data (red). (b) Low-frequency real conductivity $\sigma_{1}$ versus temperature. (O) = data. Solid line = fit to Eq.~(\ref{eqn:thermal}).}
\label{fig:Conductance_30K}
\end{figure}

The complex refractive index is then used to calculate the complex conductivity $\tilde{\sigma} = \sigma_{1} + i\sigma_{2}$, where $\sigma_{1}(\omega) = 2n\kappa\omega \epsilon_{0}$ and $\sigma_{2}(\omega) = (\epsilon_{\infty} - n^{2} + \kappa^{2})\omega \epsilon_{0}$, $\epsilon_{0}$ being the permittivity of free space, and $\epsilon_{\infty}$ the high-frequency dielectric constant. As an unknown quantity, $\epsilon_{\infty}$ was initially set to 1 when we calculate the experimental conductance, via the expression $G_{data}(\omega) = \tilde{\sigma}(\omega)t_{bulk}$; while it will be used as a fitting parameter \cite{RVAguilar2012a} in the conductance fitting for each temperature. Since the bulk state is highly insulating compared to the conducting surface states, the transport dynamics of the surface states may be modeled by a Drude (D) term, whilst the insulating bulk state by a Lorentz (L) term. Therefore, we model the conductance as $\tilde{G}_{model}(\omega) = \tilde{\sigma}_{model}(\omega) t_{bulk}$, where
\begin{equation}\label{Drude_Lorentz_All_Contributors}
\begin{split}
	\tilde{\sigma}_{model}(\omega)=& \tilde{\sigma}_{D}(\omega) f + \tilde{\sigma}_{L}(\omega) = \frac{\epsilon_{0}\omega_{pD}^{2}}{\gamma_{D} - i\omega} f \\
                     & - i\epsilon_{0}\omega(\epsilon_{\infty} - 1)
                       + \frac{\epsilon_{0} \omega_{pL}^{2}\omega}{i(\omega_{0L}^{2} - \omega^{2}) + \omega \gamma_{L}}
\end{split}
\end{equation}
with $f=t_{surf}/t_{bulk}$ being the contribution of the Drude conductivity to the model conductivity. The surface state thickness $t_{surf}$$\sim$2~nm is taken to be that of high carrier-density Bi$_{2}$Se$_{3}$ \cite{Analytis2010b}, and bulk thickness $t_{bulk}$=60~$\mu$m. The fitting parameters $\omega_{pD}$ and $\gamma_{D}$ denote the plasma frequency and scattering rate of the Drude component, respectively. $\omega_{pL}$, $\omega_{0L}$ and $\gamma_{L}$ are the plasma frequency, oscillator frequency and the scattering rate of the Lorentz component, respectively. The temperature dependence of these fitting parameters will be discussed later. Figure~\ref{fig:Conductance_30K}(a) shows the real and imaginary components of the experimental complex conductance $\tilde{G}(\omega) = G_{1}(\omega) + i G_{2}(\omega)$ of the sample at 30~K, alongside the fit to the Drude-Lorentz model, and the respective Drude and Lorentz contributions. Notice that the dips in transmission at 0.7~THz, 0.9~THz and 1.2~THz in Fig.~\ref{fig:Transmittance_All} become peaks in $G_{1}$ and $G_{2}$ at the same frequencies, in Fig.~\ref{fig:Conductance_30K}(a).

We also tried to fit our data using two Drude and one Lorentz terms, where the extra Drude term representing the free carriers in the bulk, as has been done in a THz-TDS study of Bi$_{2}$Se$_{3}$ thin films \cite{RVAguilar2012b}. We found that the Drude fitting parameters are now not well-constrained and depend sensitively on initial guesses. In Bi$_{2}$Se$_{3}$ thin films, a strong Drude component was observed at sub-THz frequencies and requires both bulk and surface Drude terms to account for it \cite{RVAguilar2012b}. The low-frequency response of our BSTS sample, on the other hand, is Lorentz-like with a Drude offset (see Fig.~\ref{fig:Conductance_30K}(a)). This is consistent with our BSTS sample having a large bulk resistivity, thus needing only one Drude term (representing the surface) in our conductance fit.

We estimate how dc conductivity of BSTS varies with temperature by plotting the low-frequency $\sigma_{1}$($\omega$=0.4~THz) versus temperature, then fitting it to the thermally-activated hopping model given by \cite{Singleton01}
\begin{equation}
\sigma(T) = A\exp{[-\Delta/k_{B}T]} + D
\label{eqn:thermal}
\end{equation} from 5~K to 150~K, where $A$ is a constant, $D$ is a non-zero offset at 0~K, $\Delta$ is the activation energy, and $k_{B}$ is the Boltzmann constant. Figure~\ref{fig:Conductance_30K}(b) shows the low-frequency conductivity together with the fit to Eq.~(\ref{eqn:thermal}), with $D = (2.80 \pm 0.01)$~($\Omega$.cm)$^{-1}$, and activation energy $\Delta$ = (26.7$\pm$0.2)~meV. The fitted value of $\Delta$ is consistent with the results by Ren \textit{et al.} that vary between 22 and 53~meV \cite{Ren11}. A similar energy scale of 30--40~meV has also been found in Bi$_{2}$Te$_{2}$Se, which has been attributed to transitions from the impurity bound states to the electronic continuum \cite{Pietro12}.

\begin{figure}\centering
\includegraphics[width=8cm,clip]{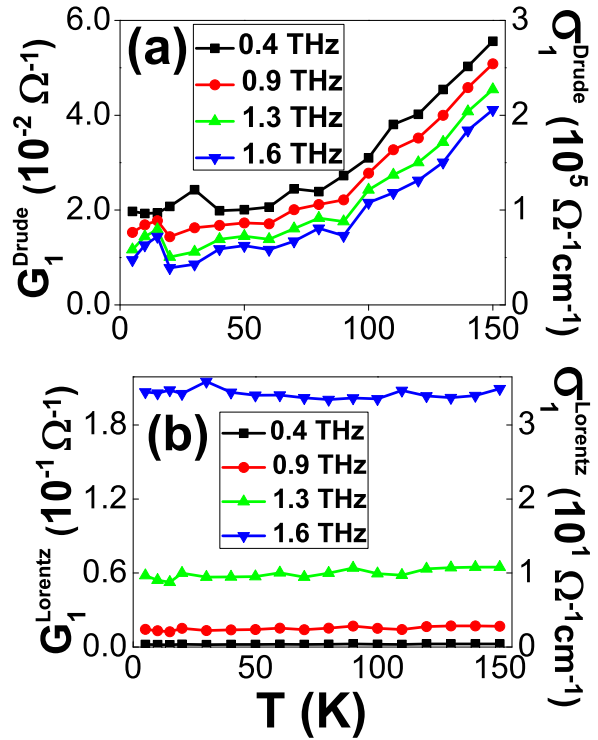}
\caption{Temperature dependence of the (a) Drude and (b) Lorentz, conductance and conductivity, at selected frequencies.}
\label{fig:Conductance_Conductivity}
\end{figure}

Figures~\ref{fig:Conductance_Conductivity}(a) and (b) show the real Drude and Lorentz conductance ($G_{1}^{\text{Drude}}$ and $G_{1}^{\text{Lorentz}}$) and conductivity ($\sigma_{1}^{\text{Drude}}$ and $\sigma_{1}^{\text{Lorentz}}$) at selected frequencies. $\sigma_{1}^{\text{Drude}}$ and $\sigma_{1}^{\text{Lorentz}}$ have been obtained by dividing each conductance component with the surface and bulk thickness respectively. At a fixed frequency, the Drude conductance increases with increasing temperature, but at a fixed temperature, it decreases slightly with increasing frequency. On the other hand, the Lorentz conductance is temperature-independent, but increases significantly with increasing frequency. The frequency dependences of the Drude and Lorentz components are just natural consequences of the respective models at low frequencies. Figure~\ref{fig:Conductance_Conductivity}(a) shows that the surface (Drude) conductivity $\sigma_{surf}\sim$10$^{5}$~($\Omega$.cm)$^{-1}$ is $\sim$10$^{4}$ times the bulk (Lorentz) conductivity $\sigma_{bulk}\sim$10~($\Omega$.cm)$^{-1}$. This suggests a significant contribution of the topological surface states to the conductivity of the BSTS sample. Note that the conductance values are independent of $t_{surf}$ and $t_{bulk}$, whereas the surface (bulk) conductivity value scales inversely as $t_{surf}$ ($t_{bulk}$). If we take $t_{surf}$ to be 40~nm as in \textit{low} carrier-density Bi$_{2}$Se$_{3}$ \cite{Analytis2010a}, then the surface-to-bulk conductivity ratio will be one order of magnitude less, which is in better agreement with electrical transport measurements \cite{Ren10}. Furthermore, if we were to replace $t_{surf}$ with $t_{bulk}$, the resulting surface-to-bulk conductivity ratio ($\sim$10$^{-1}$--10$^{0}$) would have been unphysical.

\begin{figure}\centering
\includegraphics[width=8cm,clip]{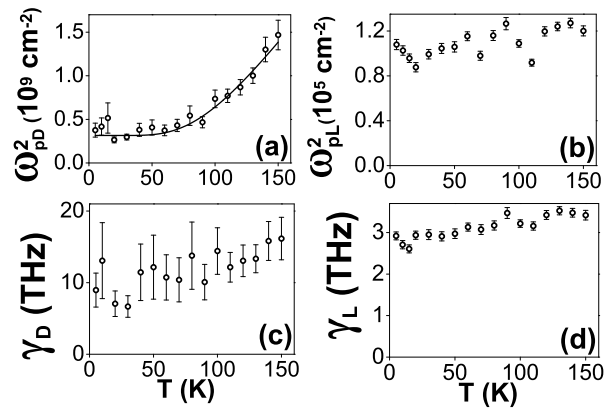}
\caption{Temperature dependence of the (a) Drude spectral weight $\omega_{pD}^{2}$, (b) Lorentz spectral weight $\omega_{pL}^{2}$, (c) Drude scattering rate $\gamma_{D}$, and (d) Lorentz scattering rate $\gamma_{L}$. Solid line in (a) is fit to Eq.~(\ref{eqn:thermal}).}
\label{fig:Spectral_Weights_Scattering_Rates}
\end{figure}

\begin{figure}\centering
\includegraphics[width=8cm,clip]{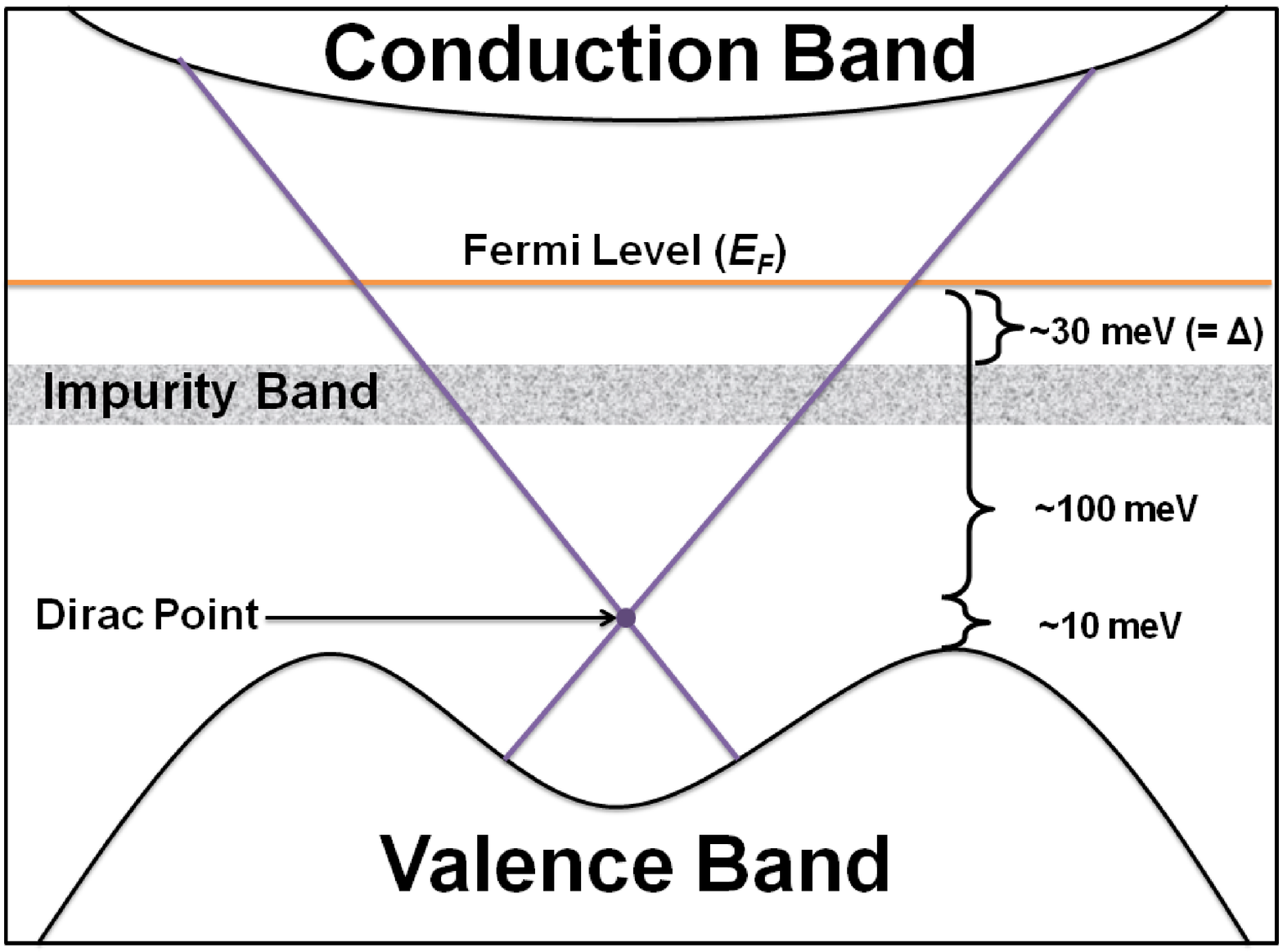}
\caption{Schematic of band diagram of BSTS, with the (donor) impurity band lying just below the Fermi level. The relative positions of the valence band, Dirac point, Fermi level and conduction band were taken from Ref.~\onlinecite{Arakane12} for a Bi$_{1.5}$Sb$_{0.5}$Te$_{1.7}$Se$_{1.3}$ sample. $\Delta$= activation gap.}
\label{fig:TI_Band}
\end{figure}

We next examine the factors that contribute to the frequency and temperature dependence of the surface and bulk conductance, by looking at the temperature dependence of the fitting parameters in Eq.~(\ref{Drude_Lorentz_All_Contributors}). Figures~\ref{fig:Spectral_Weights_Scattering_Rates}(a) and (b) display the temperature-dependent Drude spectral weight ($\omega_{pD}^{2}$) and Lorentz spectral weight ($\omega_{pL}^{2}$) respectively, while Figs.~\ref{fig:Spectral_Weights_Scattering_Rates}(c) and (d) show the surface ($\gamma_{D}$) and bulk ($\gamma_{L}$) scattering rates respectively. We found $\omega_{pD}^{2}$ to increase significantly with temperature, while $\gamma_{D}$ (proportional to charge carrier mobility) shows little change with temperature. As for bulk carrier dynamics, $\omega_{pL}^{2}$ exhibits insignificant correlation with temperature while $\gamma_{L}$ shows a slight increase with temperature. Since spectral weight is directly proportional to the charge carrier density $N$ via the relation
$\omega_{p}^{2} = N e^{2}/\epsilon_{0}m^{*}$, where $e$ denotes the elementary charge and $m^{*}$ is the effective mass of the charge carrier, this suggests the increase in the surface charge density with temperature while the scattering rate changes only marginally. This is unlike the case of conventional metals, where the increase in resistivity with increasing temperature originates from the increase in scattering rate, with the charge density remaining temperature independent. The increase in surface charge carrier density may be due to the contribution of bulk carriers from the donor impurity band to the topological surface states (Fig.~\ref{fig:TI_Band}). This band, situated $\sim$30~meV (=activation gap $\Delta$) below the Fermi level, starts to contribute carriers to the Dirac cone above 60~K, leading to a rise in the Drude spectral weight $\omega_{pD}^{2}$. This relationship between $\Delta$ and $\omega_{pD}^{2}$ is supported by our fit of $\omega_{pD}^{2}(T)$ (in Fig.~\ref{fig:Spectral_Weights_Scattering_Rates}(a)) with Eq.~(\ref{eqn:thermal}) --- a good fit was obtained with $\Delta$=(31 $\pm$ 4)~meV (solid line in Fig.~\ref{fig:Spectral_Weights_Scattering_Rates}(a)), consistent with the earlier-obtained value of $\Delta$=(26.7$\pm$0.2)~meV from the $\sigma$($\omega$=0.4~THz) fit. This is not surprising, since, from Eq.~(\ref{Drude_Lorentz_All_Contributors}), $G(\omega \rightarrow 0, T) \propto \omega_{pD}^{2}(T)$. Note that the temperature dependence of $\omega_{pD}^{2}$, and hence value of the activation gap $\Delta$, do not depend on the value of $t_{surf}$. Figure~\ref{fig:TI_Band} shows a schematic of the relative positions of the valence band, Dirac point, impurity band, chemical potential, and conduction band in BSTS. From $\omega_{pD}^{2}$, one obtains the carrier density of the Drude (surface) term $N_{surf}$$\sim$10$^{21}$cm$^{-3}$. Compare this with the carrier density of a typical metal of $\sim$10$^{23}$cm$^{-3}$ \cite{Dressel02}.

From the conductance fittings, we also obtained the oscillator frequency $\omega_{0L} = (1.92 \pm 0.04)$~THz, and temperature independent. This oscillator can be attributed to an optical phonon mode, which also agrees with the THz-TDS study by Aguilar \textit{et al.} on Bi$_{2}$Se$_{3}$ thin films \cite{RVAguilar2012b}, where a similar optical phonon has been found at $\sim$2.0~THz. This result is also consistent with the $A^{1}_{1g}$ longitudinal optical phonon obtained by pump-probe \cite{Kumar2011,Qi2010} and Raman \cite{Richter1977,Zhang2011} studies on Bi$_{2}$Se$_{3}$. A similar phonon mode has also been observed in Bi$_{2}$Te$_{3}$ thin films at 1.84~THz \cite{Wu2008}.

In conclusion, we measured the terahertz conductivity of Bi$_{1.5}$Sb$_{0.5}$Te$_{1.8}$Se$_{1.2}$ as a function of temperature using THz-TDS. By modeling the experimental conductance as a sum of a Drude term (from the surface) and a Lorentz term (from the bulk), we found the surface conductivity to be $10^{4}$ times greater than the bulk conductivity. The high surface conductivity compared to the bulk suggests the presence of metallic surface states of BSTS.

L.W. (E.E.M.C.) acknowledges funding from Singapore National Research Foundation RCA-08/018 (NRF-CRP4-2008-04) and Singapore Ministry of Education AcRF Tier 2 MOE2010-T2-2-059 (ARC 23/08). J.-X.Z. is supported by the National Nuclear Security Administration of the U.S. DOE at LANL under Contract No. DE-AC52-06NA25396, the U.S. DOE Office of Basic Energy Sciences, and the LDRD Program at LANL.

\bibliography{BiSbTeSe}

\end{document}